\documentstyle[preprint,aps,prb,epsf]{revtex}

\begin{document}

\title{Temperature-dependent electron paramagnetic resonance studies of
charge ordered Nd$_{0.5}$Ca$_{0.5}$MnO$_3$ }
\author{ Janhavi P. Joshi, Rajeev Gupta,  A. K. Sood and S. V. Bhat}
\address{
Department of Physics,
Indian Institute of Science,
Bangalore 560 012, India
}
\author{ A. R. Raju and C. N. R. Rao$^\dagger$}
\address{
$^\dagger$CSIR Centre for Excellence in Chemistry, Jawaharlal Nehru Centre for Advanced
Scientific Research, Jakkur P.O, Bangalore 560 064, India}
\maketitle
\widetext
\begin{abstract}
We report  electron paramagnetic resonance measurements  on  single crystalline and powder samples of Nd$_{0.5}$Ca$_{0.5}$MnO$_{3}$ across the charge-ordering transition at $T_{co}=240 K$ down to the antiferromagnetic ordering transition at $T_N = 140 K$. The changes in the linewidth, g-factor and intensity as  functions of temperature are studied to understand the nature of spin-dynamics in the system. We explain the observed large decrease in the linewidth from $T_N$ to $T_{co}$ in terms of motional narrowing caused by the hopping of the Jahn-Teller polarons yielding an activation energy of $E_a = 0.1$ eV. Similar analysis of data on Pr$_{0.6}$Ca$_{0.4}$MnO$_3$ published earlier gives $E_a=0.2$ eV.  Below $T_{co}$, the g-value increases continuously suggesting a gradual strengthening of the orbital ordering. We  give a qualitative explanation of  the maximum in the asymmetry ratio A/B observed at $T_{co}$ and its temperature dependence in single crystal spectra which also supports the model of motional narrowing.  


\end{abstract}

\section{Introduction}

Doped perovskite manganites of the form RE$_{1-x}$A$_x$MnO$_3$ where RE is a trivalent rare earth ion such as La$^{3+}$, Pr$^{3+}$ and Nd$^{3+}$, and A is a divalent alkaline earth ion such as Ca$^{2+}$, Sr$^{2+}$ and Ba$^{2+}$ are mixed valent systems containing Mn$^{3+}$ and Mn$^{4+}$. They exhibit a multitude of magnetic, electronic and structural phase transitions as  functions of doping level x (which controls the Mn$^{3+}$ to Mn$^{4+}$ ratio),
 temperature, magnetic and electric fields\cite{yoo,cnr,toku}. The interplay of charge, spin and orbital degrees of freedom in these systems results in a substantial fragility of the phase boundaries with respect to the varying physical parameters. The dependence of physical properties on the choice of RE and A and their sizes can be quantitatively understood in terms of the tolerance factor t, defined as ${t = {<r_{RE,A>} + r_o \over {\sqrt{2} (<r_{Mn}> + r_o)}}}$, where $<r_{RE,A}>$ is the average ionic radius of the  rare earth or the alkaline earth ion, $<r_{Mn}>$ is the average ionic radius of the manganese ions and $r_o$ is the  oxygen ion radius. For $x=0.5$ and a certain range of $t (> {\sim 0.975})$,\cite{kuwa} these systems exhibit the much studied phenomenon of colossal magnetoresistance (CMR). CMR refers to the large negative   change in the resistivity of the material on the application of a magnetic field. In zero field these systems show  an insulator to metal transition coincident with a paramagnetic to ferromagnetic transition implicating the connection between the electronic and spin degrees of freedom. For $0.975<t<0.992$,  the ferromagnetic metallic state becomes unstable with respect to an insulating, antiferromagnetic, charge ordered (CO) state (e.g. in Nd$_{0.5}$Sr$_{0.5}$ MnO$_3$) below a certain temperature. The CO state consists of real space ordering of Mn$^{3+}$ and Mn$^{4+}$ ions in the material, a phenomenon similar to Wigner crystallization \cite{rao2}. Further, for 
$t\le 0.975$ and $0.3\le x\le$0.7  as in Pr$_{0.6}$Ca$_{0.4}$MnO$_3$ and Nd$_{0.5}$Ca$_{0.5}$MnO$_3$ only a transition  to a CO state is observed on cooling while the material becomes antiferromagnetic at a further lower temperature.

The metallic ferromagnetic ground state of the manganites is understood in terms of  Zener's double exchange(DE) model\cite{zene,ande,genn}. The basic feature of DE is the hopping of a d-hole from Mn$^{4+}$ to Mn$^{3+}$ via the oxygen which can also be looked upon as  the transfer of an electron from the Mn$^{3+}$ site to the central oxygen ion and simultaneously the transfer of an electron from the oxygen ion to the Mn$^{4+}$ ion. Since such a transfer is most probable when the spins of the $t_{2g}$ electrons of  the Mn$^{3+}$ ion are aligned with the $t_{2g}$ spins of the adjacent  Mn$^{4+}$ ion, ferromagnetism occurs concommitantly with metallic conduction. Mn$^{3+}$ ions being strong Jahn-Teller (J-T) ions, the mobile $e_g$ electron is also expected to carry the lattice distortion with it making the polaronic contribution to the conduction an important factor as well\cite{mill}. As far as the CO phenomenon is concerned, one of the possible origins of it is thought to be the strong  
intersite electronic repulsive interaction normally present in the transition metal based oxides\cite{mishra}. However the long range Coulomb repulsion alone cannot explain the observed high sensitivity of the CO state to an applied magnetic field because of which the CO state of some systems ``melts" into a ferromagnetic metallic state. This result points towards a role for the spins of the carriers as well.

Since electron paramagnetic resonance (EPR) is a powerful probe of spin dynamics, a number of EPR studies have been performed on CMR manganites aiming at understanding the microscopic nature of the interplay between spin and charge degrees of freedom\cite{see,hube,hube1,osef,riva,caus,tova,shen,lof1,ivan,shen1}. EPR results on the CMR materials show some characteristic features. The linewidths ($\Delta$H) are large and show a minimum around the ferromagnetic transition temperature $T_c$, increasing as a function of temperature on either side of it.  Considerable amount of controversy exists regarding the interpretation of the $\Delta$H dependence on T for $T > T_c$. Seehra {\it et al.,}\cite{see} in an early study attributed this behaviour to spin-phonon interaction. While this interpretation was questioned in the later reports by other workers\cite{hube,caus}, present consensus seems to be that the linewidths have contributions from  two main  interactions, J-T distortion mediated crystal-field interactions(CF)  and  anisotropic Dzyaloshinsky-Moriya(DM) exchange interaction.  The temperature dependence of the EPR linewidths based on  these interactions has been calculated\cite{hube,caus} and the results seem to match the experimental findings quite well. However, Shengelaya {\it et al.,}\cite{shen1}  noticed a close similarity between the temperature dependent increase in the EPR linewidths and the conductivity in these materials  and proposed a model based on the hopping of small polarons.   The activation energy obtained from the linewidth dependence on temperature turns out to be similar to that obtained from the conductivity measurements. Ivanshin {\it et al.,}\cite{ivan} indicate that different mechanisms may be operative in different regimes of x and lend support to the model proposed in ref. 10 for $0.075 \le x \le 0.15$.

 In contrast,  the only published EPR work on a charge-ordered manganite to date is that on Pr$_{0.6}$Ca$_{0.4}$MnO$_3$(PCMO)\cite{gupta} . In this work it was found that below the charge ordering transition temperature $T_{co}$ the linewidth increased slowly with the decreasing temperature (apart from a significant jump at $T_{co}$) before saturating at temperatures close to $T_N$. On the high temperature side of $T_{co}$, the temperature dependence was much weaker over the relatively small temperature range that was covered. In this study from the temperature dependence of the intensity above $T_{co}$, the ferromagnetic exchange coupling constant was calculated to be 150 K. Further, the EPR g-factor showed the following  interesting behaviour: (1) A g shift opposite to the one expected for Mn$^{3+}$ and Mn$^{4+}$ was observed. (2) Below $T_{co}$ a gradual increase of g was observed with decreasing temperature, which was interpreted to be a signature of gradual strengthening of orbital ordering. (3) It was noted that the magnitude and the behaviour of g were different from those reported for the CMR manganites where a temperature independent $g \sim 2$ was observed.

In this work we report the EPR study of Nd$_{0.5}$Ca$_{0.5}$MnO$_3$ (NCMO) in the temperature range 4.2 to 300 K covering the antiferromagnetic ordering temperature $T_N$ and the charge ordering temperature $T_{co}$.  At zero field, NCMO with $t=0.930$ is an insulator throughout the temperature range with $T_{co}=240K$ and $T_N=140K$. Below $T_N$ an antiferromagnetic phase with complete charge ordering and orbital ordering is observed. Between $T_N$ and $T_{co}$,  the orbital ordering gradually develops as the temperature is lowered from $T_{co}$ to $T_N$. At low fields both the antiferromagnetic phase and the CO phase have small magnetic susceptibility. At higher fields($>10$~T)\cite{millange}, however, a spin-flip transition occurs and the ordering becomes ferromagnetic and the charge ordered state melts.  In the present work we offer  an  explanation for the temperature dependence of the EPR linewidths in charge ordered manganites including NCMO and PCMO, in terms of ``motional narrowing" which we believe is particularly applicable to the behaviour between $T_N$ and $T_{co}$. From a qualitative understanding of the temperature dependence of the asymmetry ratio A/B, including the maximum observed at $T_{co}$, we obtain an order of magnitude estimate of the electron diffusion time and show that it is consistent with the picture of ``motional narrowing". The similarity between the experimental results of PCMO and NCMO shows that the observed features are fingerprints of the CO state.

\section{ Experimental Details}

The single crystals of NCMO were prepared by the float zone technique. The dc magnetic susceptibility shows  a large peak at $T_{co}= 240 K$ and a relatively smaller peak at $T_N = 140 K$\cite{millange,rao3,rao}.  The resistivity which is weakly dependent on temperature for $T > T_{co}$ shows a strong temperature dependence below $T_{co}$, increasing by nearly three orders of magnitude from $T_{co}$ to $T_N$ \cite{millange,rao3,rao}. The EPR experiments were carried out on both single crystal and powder samples using a Bruker X band spectrometer (model 200D) equipped with an Oxford Instruments continuous flow cryostat (model ESR 900). The spectrometer was modified by connecting the X and Y inputs of the chart recorder to a 12 bit A/D converter which in turn is connected  to a PC enabling digital data acquisition. With this accessory, for the scanwidth typically used for our experiments i.e 6000 Gauss, one could determine the magnetic field to a precision of $\sim$ 3 Gauss.For single crystal study the static magnetic field was kept parallel to the c-axis of the crystal. The temperature was varied from 4.2 K to room temperature (accuracy: $\pm 1 K$) and the EPR spectra were recorded while warming the sample. For measurements on powder, the powder was dispersed in paraffin wax. While doing experiments on both the single crystal and the powder, a speck of DPPH marker was used to ensure the accurate determination of the g-value of the sample.

\section{Results and Discussion}

Figures 1a and 1b show the EPR spectra (${dP \over dH}~{\it vs}~H $) recorded in the temperature range 290 K to 180 K for single crystal and powder samples respectively. Below 180 K the signals were too weak to be analyzed and below $T_N$, no signal was observed.  In these signals the sharp signal due to DPPH, used as the field marker has been digitally subtracted to aid the fitting of the lineshapes. As can be seen, the lineshapes in the two cases differ significantly. In single crystals we observe a characteristic Dysonian lineshape (${A\over B} > 1$, where A and B are the amplitudes of the low field and high field halves of the signal, respectively)  while in  the powder sample a symmetric Lorentzian line is observed. The asymmetric Dysonian lineshapes result from a mixture of the absorptive and dispersive components of the susceptibility, caused by the non uniform distribution of the microwave electromagnetic field due to the sample size being larger than the skin depth\cite{dyso,feh}. Along with this the motion of the paramagnetic centres can also contribute to this asymmetry. Since the lines are very broad both in powder and single crystals, for accurate determination of the various lineshape parameters we have fitted the signals to appropriate lineshape functions. For the single crystal spectra we used  the  equation\cite {ivan}
$$ {dP\over dH} = {d\over dH}({\Delta H+\alpha (H-H_0) \over (H-H_0)^2 + \Delta H^2}+ {\Delta H+\alpha (H+H_0) \over (H+H_0)^2 + \Delta H^2})\eqno(1) $$
where $H_0$ is the resonance field, $\alpha $ is the fraction of the dispersion component added into the absorption signal and  $\Delta H $ is the line width. 
  The use of the two terms in the equation accounting for the clockwise as well anticlockwise circularly polarised component of microwave radiation is necessary because of the large width of the signals. 

The symmetric powder signals (Fig. 1(b)) are fitted to the Lorentzian shape function also incorporating the two terms as follows: 
$${dP\over dH} = {d\over dH}({\Delta H \over (H-H_0)^2 + \Delta H^2}+ {\Delta H \over (H+H_0)^2 + \Delta H^2})\eqno (2) $$

As can be seen from Fig. 1, the fits of the signals to the two lineshape functions are excellent. The fitting parameters thus obtained are plotted as  functions of temperature in Figs. 2 and 4. Figure 3 shows the temperature dependence of the A/B ratio (defined in the inset), obtained from the fitted lineshapes. The g-values have been obtained from the fitted centre field values $H_0$, taking $g=2.0036$ for DPPH. The linewidths plotted are peak to peak linewidths calculated from the Lorentzian full widths at half maxima(FWHM) obtained from the fits using
$\Delta H_{pp} = {\Delta H_{FWHM} \over \sqrt 3} $

The origin of the EPR signal in manganites has been the subject of some discussion in literature. Normally, Mn$^{3+}$ (3d$^{4}$, S=2) EPR is  difficult to observe because of the large zero-field splitting and strong spin-lattice relaxation. However, a tetragonal J-T distortion makes it observable\cite{shen}. It was recognised that the signals in manganites cannot be due to isolated  Mn$^{4+}$ (3d$^{3}$, S=3/2) ions alone and all the Mn ions present i.e of both Mn$^{3+}$ and Mn$^{4+}$ types were concluded to contribute to the signals \cite{}. The EPR intensity is expected to be proportional to the dc susceptibility $\chi_{dc}$ of the spins.   This is borne out by  the inset of Fig. 4c, where we show the product of the dc magnetisation M and temperature T plotted as a function of T  (adapted from ref. 24). Two peaks are seen in  M x T {\it vs} T curve, a large one at $T_{co} $= 250 K and a smaller  one at T$_N =$ 140 K.  Interestingly I$_{EPR}$ x T {\it vs} T for the powder sample shown in fig 4c is seen to follow M x T {\it vs} T closely indicating the proportionality  between $\chi _{dc}$ and I$_{EPR}$.

The temperature dependence of the asymmetry parameter A/B is shown in Fig. 3. The insets of the figure indicate the procedure adopted to determine the ratio A/B. It is clear that one needs to determine the baseline of the signal accurately to obtain an accurate value of A/B. However because of the large width of the signals, it was not possible to experimentally determine the baseline. Therefore, the fitted signal was extended to high values of magnetic field($\sim $ 30,000 Gauss) till a {\it nearly } horizontal baseline was obtained. Ideally one should observe the baseline on the low field side at the same level as that on the high field side . However, occasionally EPR signals, especially of the Dysonian lineshapes \cite{kodera} exhibit a mismatch between the low field and the high field baselines. Therefore we have joined the high field baseline, obtained from extrapolation, to the zero field value of the fitted signal to determine the overall signal baseline and to calculate the A/B ratio. Obviously this procedure leads to some error in the values of the latter. However, the fact that the trend of the temperature dependence of the ratio including its maximum is correctly reproduced  can be seen from the two insets to Fig. 3, one for 225K and  another for 190K. We have also performed an independent experiment with a thicker sample and verified that the values presented in Fig. 3 are actually lower than those for the thicker sample, thus rendering credence to the arguments to follow. From the plot of A/B {\it vs} T shown in Fig. 3 it can be seen that, starting from room temperature to close to $T_{co}$ the A/B ratio remains essentially constant at a value $\sim$ 2.75. This value, being higher than 2.55 expected for stationary spins\cite{kodera} indicates that the paramagnetic centres are mobile. At $T_{co}$ it undergoes a discontinuous increase to $\sim$ 4. Further cooling results in a continuous decrease as expected from the monotonic increase in the resistivity of the sample.  Similar but sharper change in A/B consistent with the sharper jump in resistivity was also observed at $T_{co}$ in PCMO\cite{gupta}. A qualitative understanding of this behaviour can be obtained by taking into account the subtleties of  the Dyson effect. As discussed by Kodera\cite{kodera}, the A/B ratio depends in a complex manner on various material parameters such as the ratio $\lambda$ of the sample thickness $\theta$ to the skin depth $\delta$, electron diffusion time through the skin depth $T_D$ and the spin-spin relaxation time $T_2$. For certain ranges of these parameter values, as shown by him, A/B can go through a maximum. (Figs. 8 and 10 of reference 27 ). In NCMO and PCMO, the transition to the CO state results in values of $\delta$ (through the changes in $\rho$), which along with the values of $T_D$ and $T_2$ make the  A/B go through a maximum.   Referring again to the analysis by Kodera, a peak value of A/B of $\sim$ 4 with $\lambda$ in the range of 2 to 3,(which is reasonable for our sample size of $\sim $ 1 mm, and $\rho $ of $\sim $1 $\Omega $-cm \cite {rao} just below T$_{co}$)  implies a value in the range of 1 to 5 for $(T_D/T_2)^{1/2}$(Fig. 5 of reference 27) where $T_2={2\over\sqrt3}{h\over g \beta \Delta H_{pp}}$. It is well known that when the motional frequencies are comparable to the strength of the broadening interactions (expressed in frequency units), ``motional narrowing" of the linewidth occurs. Thus the fact that $T_D$ is of the same order of magnitude as $T_2$ provides additional support to the model of ``motional narrowing" to be discussed next.

Figures 2a and 4a show the temperature dependence of the linewidth  in the single crystal and the powder samples respectively. It is noted that starting from room temperature down to $T_{co}$ the linewidth decreases very slowly with temperature below which it increases with decreasing temperature, by a factor of two over the temperature range from 230 K to 160 K. We note that this increase in  the linewidth is different from the  behavior in CMR manganites. The $\Delta H $(T) in the latter has been the subject of some controversy in the literature \cite{riva,dom,riva1}. While in the ceramic and thin film samples $\Delta H $ diverged after reaching a minimum at $T_{min}$ ($\sim$ 1.1 $T_c$ where $T_C$ is the ferromagnetic transition temperature.), in as grown single crystal samples $\Delta H $ remained independent of T below $T_c$. The same exhibited an increase in $\Delta H $ with decreasing T when the surface was polished to create craters of size 3 $\sim $ 8 $\mu m$. Dominguez {\it et al.,}\cite{dom} attributed the increase in $\Delta H $ below $T_c$ in ceramic and thin film samples to chemical and magnetic inhomogeneities. Rivadulla {\it et al.,}\cite{riva,riva1} showed that the demagnetisation fields arising from pores in polycrystalline samples  and surface polished single crystals \cite{} are responsible for the increase in $\Delta H $. The systems studied by these authors differ from our samples in one important respect. They are in the long range ferromagnetically ordered state whereas we are concerned with the charge ordered state. Indeed it was found\cite{riva1} that $\Delta H $(T) for T $>$ T$_{min}$ was proportional to magnetisation M(T) in these materials whereas in our systems, while $\Delta H$ increases with decreasing T, the magnetisation shows a non-monotonic behaviour, {\it decreasing} with decreasing T for most of the temperature range $T_N < T < T_{co}$ \cite{}.

Two questions are interesting in this context: (1) What is the origin of the linewidth? (2) What is the mechanism that narrows down the signal while going from $T_N$ to $T_{co}$? Huber \cite{hube1} argues that in CMR manganites for $T > T_c$, the exchange narrowed dipolar linewidths must be orders of magnitude smaller than the observed values and therefore the dipolar interaction cannot be the cause of  the linewidths. The magnitude and the temperature dependence of $\Delta H $ then could  be qualitatively explained with the assumption that the linewidth arises due to the anisotropic crystal-field(CF) effects and the Dzyloshinsky-Moriya(DM) exchange interactions.  While it is likely that for $T > T_{co}$ in NCMO and other CO manganites, a mechanism similar to that observed for $T > T_c$ in CMR manganites is operative, it is clearly different for $T < T_{co}$ since the T dependence is quite the opposite. Moreover, the alternate arrangement of Mn$^{3+}$ and Mn$^{4+}$ ions obtained in the CO state could lead to ``exchange broadening" due to hetero-spin dipolar interaction instead of the ``exchange narrowing" observed for homo-spin dipolar interaction. \cite{abra} Keeping in mind the fact that the CO state culminates into an antiferromagnetically ordered state at $T_N$, we now compare our results with EPR results of other antiferromagnetic materials in their paramagnetic state (i.e for $ T > T_N$). A number of such studies have been reported starting with the early work of Burgiel and Strandberg \cite{burg} on Mn$F_2$ to the more recent work on CuO by Monod {\it et al.,}\cite{monod}. Both three dimensional pseudocubic antiferromagnets (AFs) such as RbMnF$_3$ and two dimensional AFs such as K$_2$MnF$_4$ have been studied \cite{gupta1,gupta2,gul,hube2,rich,van}. A common feature of EPR in all these materials is that approaching $T_N$ from above $\Delta H $ gradually decreases till close to $T_N$ where it quite sharply diverges. Thus, quite interestingly in the paramagentic phases of both antiferromagnetic and ferromagnetic systems the EPR linewidth {\it decreases} as the temperature is decreased towards the transition temperature. Our results on NCMO and on the previously reported PCMO show that, the behaviour in CO systems is exactly opposite; $\Delta H $ decreasing as the temperature is increased above $T_N$.  In the same temperature range, the resistivity also decreases due to the activated hopping of the charge carriers viz. the Jahn-Teller polarons.  The hopping motion of these Jahn-Teller polarons involves the hopping of $e_g$ electrons with its associated spin from one site (Mn$^{3+}$) to another site (Mn$^{4+}$). This random motion of the magnetic moments can lead to ``motional narrowing" of the linewidth as suggested by Huber \cite{hube1} in a slightly different context. 

 An analogy can be drawn between this situation and the motion of the ions in fast ionic conductors where the NMR linewidth which is the result of intermolecular dipolar interaction decreasing with increasing temperature due to an increase in ionic conductivity. This is a result of the ``motional narrowing" of the NMR linewidths.  We believe that the narrowing of the EPR signals in the CO manganites can be understood along similar lines, the hopping of the $e_g$ electrons leading to the averaging out of the interactions between the Mn$^{3+}$ and Mn$^{4+}$ magnetic moments such as the DM interaction.  The motion can also decrease the effect of the crystal field distortion on the linewidth. 

In the discussion of ``motional narrowing" in NMR, the fluctuations which have significant spectral density around the frequency corresponding to the strength of the broadening  interaction are known to have the maximum effect in averaging out the interaction. Assuming an exponential decay of the corresponding correlation function a semi empirical formula\cite{abra1} 
$${\delta \omega ^2 }= \delta \omega_{0} ^{"^2} + \delta \omega_{0} ^{'^2}\* {2\over \pi}\*tan^{-1}(\alpha \delta \omega \tau_c)\eqno(3) $$
where $\delta \omega$ is the linewidth of the signal, $\delta \omega_{0} ^{"}$ is the residual linewidth, $\delta \omega_{0} ^{'}$ is the rigid lattice linewidth, $\alpha$ is a factor of the order of unity and $\tau_c$ is the correlation time,
is used to describe the process of linewidth decrease with increase in temperature and to extract the corresponding correlation times. We have carried out similar exercise in the analysis of the linewidths of NCMO and PCMO single crystal data (data taken from ref.22). While qualitatively the ``motional narrowing" is a reasonable explanation for the temperature dependence of the linewidth between $T_N$ and $T_{co}$, one is faced with some problems in the quantitative analysis of the same. Because, as we see from Fig. 2a, the linewidth has not reached its `rigid lattice' value, the process being pre-empted  by the occurrence of the transition to the antiferromagnetic state. We have therefore taken the largest width just above $T_N$ as the `rigid lattice' linewidth $\delta \omega_{0} ^{'}$ and the smallest width below $T_{co}$ as the residual width $\delta \omega_{0} ^{"}$. Thus the rigid lattice linewidth and the residual linewidths are taken to be 3124 and 1208 Gauss, respectively for NCMO and 2773 and 1587 Gauss, respectively for PCMO. 

 In Figs. 5a and 5b we present the results of $\tau _c$ dependence on temperature for NCMO and PCMO single crystals. Assuming an Arrhenius dependence of $\tau _c$ on T of the form $\tau_c = \tau_0e^{Ea\over k_BT}$, where $k_B$ is the Boltzmann constant,  we estimate the activation energy $E_a$ to be 0.1 eV and 0.2 eV  for NCMO and PCMO, respectively which are close to the values obtained from other experiments. For example, Vogt {\it et al.,} \cite {rao} obtain $E_a=0.12$~eV from $\rho$-T measurements on NCMO. Similarly a value of 0.2 eV is obtained for the $E_a$ of PCMO \cite {guha}. In view of the approximations made regarding the rigid lattice and residual linewidths, our values of $E_a$ should be taken only as approximate. By varying the two linewidths by about 5 percent, we find that $E_a$ also changes by about 10 percent. Even then, the fact that our values are of similar magnitudes as those obtained from other experiments points towards the essential correctness of the approach.

Figure 4b shows the temperature dependence of the g-factor in the powder sample. The behaviour closely follows that observed in PCMO earlier by us. Both the unexpected positive g shift and an increase in the g value as the temperature is decreased are observed in NCMO as well. Since in the powder sample it is  expected that the internal field effects are averaged out, we believe that the observed variation of g with temperature is  intrinsic to the sample. This can possibly be explained by the changes in the spin-orbit coupling constant consequent to the  orbital ordering. The effective g-value for a paramagnetic centre is given by $g_{eff} = {g(1\pm {k\over \Delta})}$ where $\Delta$ is the crystal field splitting and k is the spin-orbit coupling constant.  The gradual build up of orbital ordering taking place  when the temperature is decreased from $T_{co}$ to $T_N$ can change the spin-orbit coupling as well as the crystal field splitting which can give rise to the observed increase in the g-value. 

As mentioned in section I, in manganites charge, spin, lattice and orbital degrees of freedom are intercoupled and the result of any experimental measurement may reflect contributions from more than one of these parameters. For example, it may be possible that the changing nature of the magnetic fluctuations i.e., from antiferromagnetic to ferromagnetic as the temperature is varied from $T_N$ to $T_{CO}$ could lead to the observed decrease in $\Delta H$ and g. However we note that while $\Delta H$ and g decrease monotonically with increasing T in a manner analogous to the behaviour of resistivity, magnetisation shows a non-monotonic behaviour. Further, the lattice constants of the crystal are shown \cite{millange} to change continuously from $T_N$ to $T_{CO}$ such that the distortion of the oxygen octahedra continuously changes. This would lead to a continuous change in the crystal field and therefore in the g value. Our conclusions related to $\Delta H$(T) and g(T) should be viewed in the light of this discussion.

Now we consider the effects of possible phase segregation in the sample on the temperature dependence of $\Delta H $ and g because it is conceivable that such phase separation can lead to the increase in $\Delta H $ and g with decrease in T.
Manganites are known to exhibit submicroscale coexistance of two competing phases, one, a hole- rich ferromagnetic phase and another, a hole-poor antiferromagnetic CO state. For example Liu { \it et al.,} \cite{liu,bao} interpret the results of their optical reflectivity study on Bi$_{1-x}$Ca$_x$MnO$_3$ (x $\ge$ 0.5), as signifying the phase separation behaviour in which domains of antiferromagnetic and ferromagnetic order coexist. Uehara { \it et al.,} \cite{ueh1} provide electron microscopic evidence for phase separation of (La,Pr,Ca)MnO$_3$ into a mixture of insulating  and metallic ferromagnetic regions. However, as amply illustrated in the recent review article by E. Dagotto { \it et al.,} \cite{dag} as yet there is no clear understanding of the cause or  nature of the phase separation. In fact, there is some experimental evidence against phase separation. For example Mukhin { \it et al.,}\cite{mukhin} interpret the results of antiferromagnetic resonance experiments in La$_{1-x}$Sr$_x$MnO$_3$ as evidence against electronic phase separation. Therefore, since the possibility of occurence of phase separation sensitively depends on the actual system, the nature of the phase transition, the level of doping and the rate of cooling \cite{ueh},it is necessary to examine the actual system being studied from this point of view. NCMO has recently been carefully studied by Millange  { \it et al.,}\cite{millange}
 by neutron diffraction and they find no evidence of any mixed phases for $T_N < T < T_{co}$. Instead, they find, as the temperature is decreased from $T_{co}$ to $T_N$, ferromagnetic correlations continuously decrease while the antiferromagnetic correlations increase. Based upon this result, we feel that the behaviour of $\Delta H $ and g in NCMO is not a consequence of phase separation but can be attributed to the charge ordering at $T_{co}$ and the gradual development of orbital ordering as the sample is cooled from $T_{co}$ to $T_N$. However further controlled experiments and calculations may be necessary to come to a definite conclusion about this aspect. 

\section{Summary}

In summary we report  EPR measurements on the charge ordering manganite Nd$_{0.5}$Ca$_{0.5}$MnO$_3$. We observe that various parameters of the EPR signals like linewidth,  intensity, asymmetry parameter and g-value are sensitive functions of temperature  and these parameters also mark the charge ordering transition in this material. The observed change in the linewidth in the temperature range below $T_{co}$ can be explained using the semiempirical model of ``motional narrowing".  The magnitude and the temperature dependence of the asymmetry ratio A/B support this model. Assuming an Arrhenius dependence of correlation time we estimate the activation energy of electron hopping to be 0.1 eV for NCMO and 0.2 eV for PCMO which are consistent with the results of other measurements.  The g variation below $T_{co}$ possibly tracks the gradual strengthening of the orbital ordering and increasing crystal field effects. 

\section{Acknowledgments}

The authors acknowledge the help of Sachin Parashar in sample preparation. 
JPJ would like to thank  CSIR, India for a senior research fellowship.
SVB and AKS thank the Department of Science and Technology for financial support.

\newpage

\begin{large}
Figure Captions
\end{large}

\noindent { FIGURE 1:}
EPR spectra of (a) single crystal and (b) powder sample of Nd$_{0.5}$Ca$_{0.5}$MnO$_3$ for a few temperatures. The signal from DPPH has been subtracted. The solid line shows the fit of the experimental data to Eqs. 1 and 2 for (a) and (b), respectively.\\

\noindent { FIGURE 2:}
Temperature variation of the lineshape parameters for the single crystal sample; (a) peak to peak linewidth $\Delta H_{pp}$ and (b)~g-factor \\ 

\noindent { FIGURE 3:}
Variation of A/B ratio with temperature in single crystal spectra. The insets illustrate the method adopted to calculate the A/B ratio. EPR signals at two different temperatures (225K and 190K)  (filled circles) with different A/B ratios, fitted to the Dysonian line shape of Eq.1 (the solid line) are shown. The fitted signal is extended to a high field ($\sim$ 20000 and $\sim$ 30000 Gauss resp.) to obtain the base line.\\

\noindent { FIGURE 4:}
Temperature variation of the Lorentzian lineshape parameters for the powder sample. (a) peak to peak linewidth (b) g-factor (c) intensity times the temperature. The inset of (c) shows the product of magnetisation M for H$\parallel$c and temperature T plotted as a function of T (adapted from ref. 24)\\

\noindent { FIGURE 5:}
ln $\tau_c$ {\it vs.} 1/T for (a) Nd$_{0.5}$Ca$_{0.5}$MnO$_3$  and (b) Pr$_{0.6}$Ca$_{0.4}$MnO$_3$,  obtained from Eq. 3. The solid lines are fits to the Arrhenius equation.

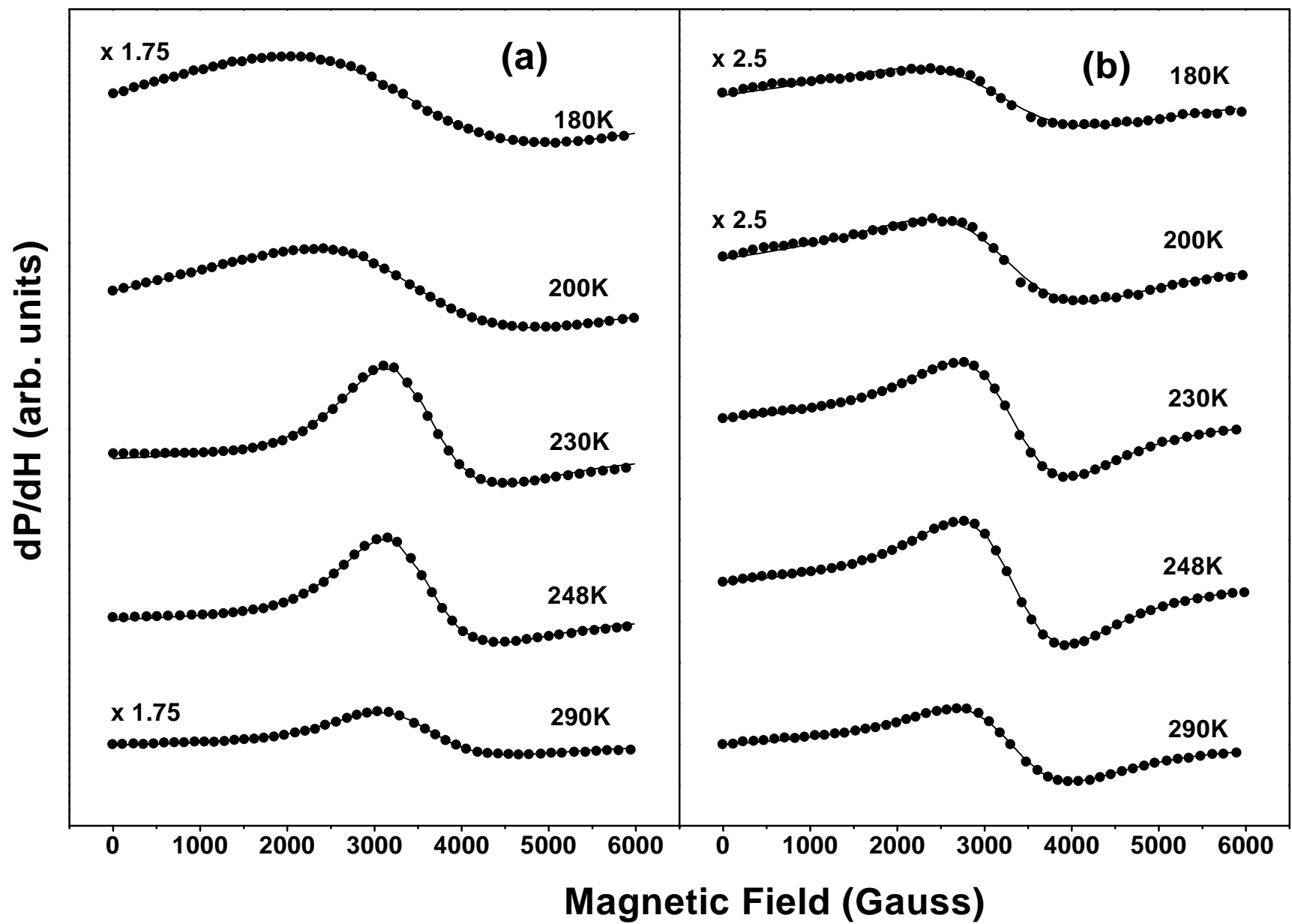
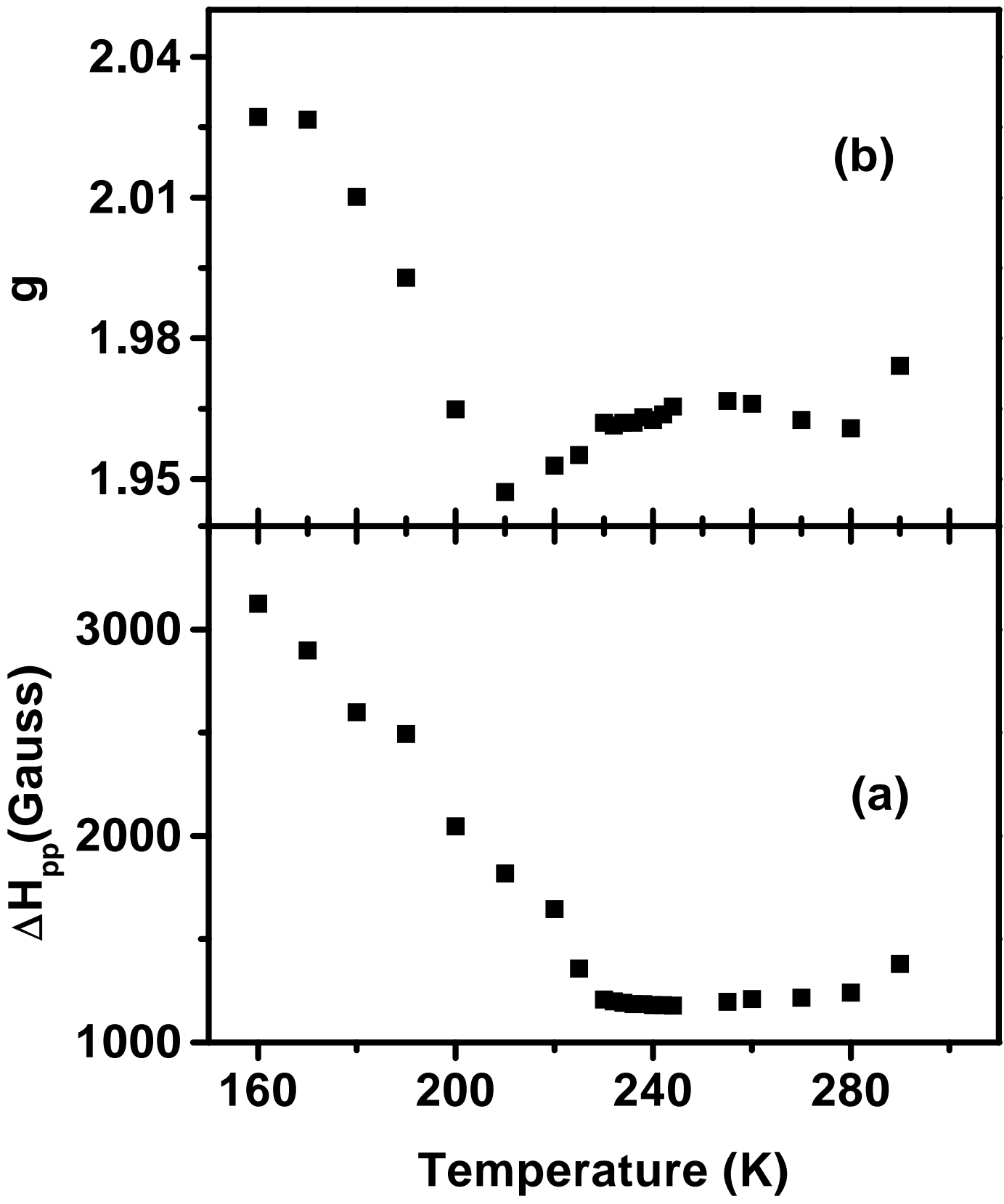
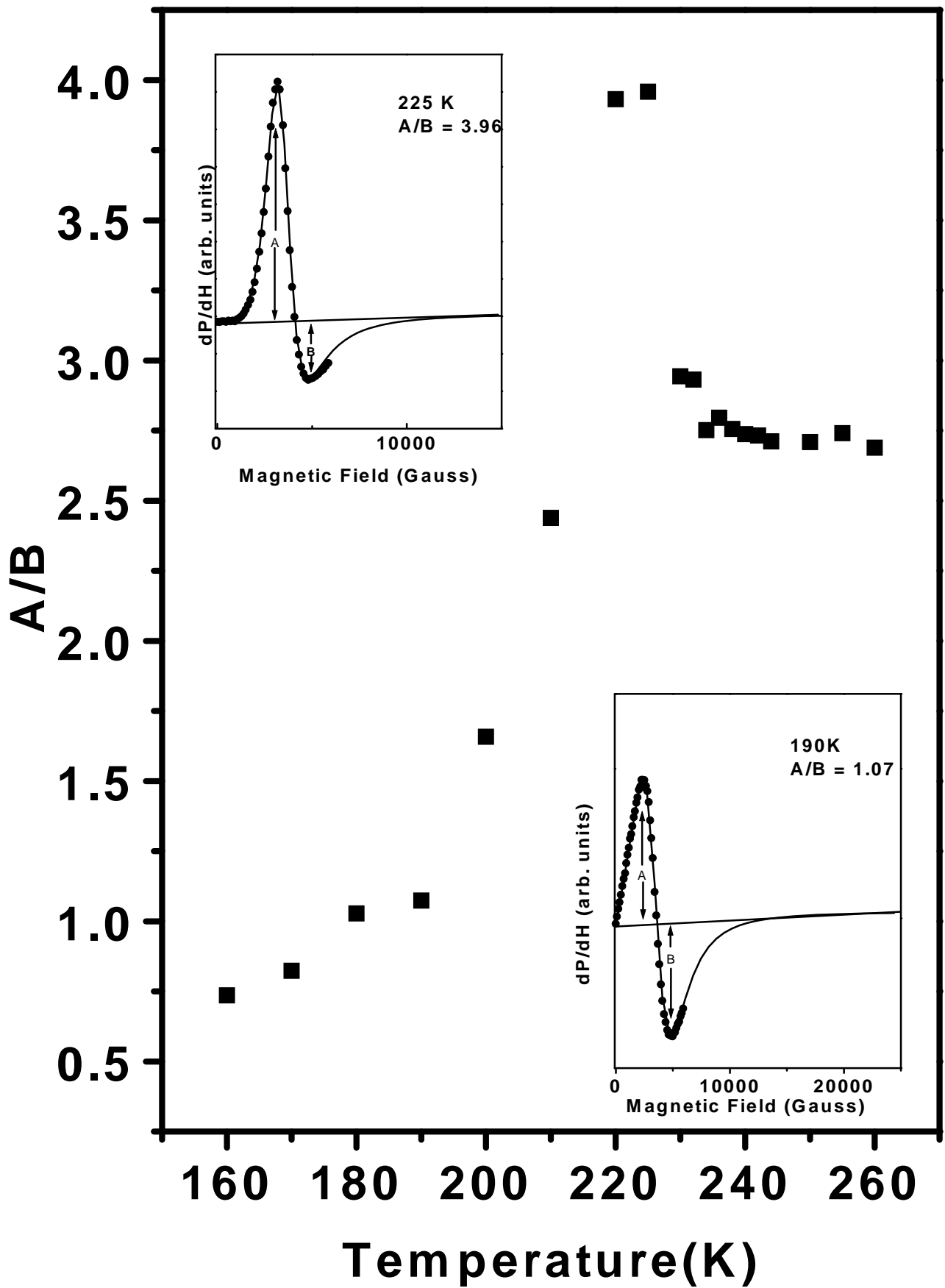
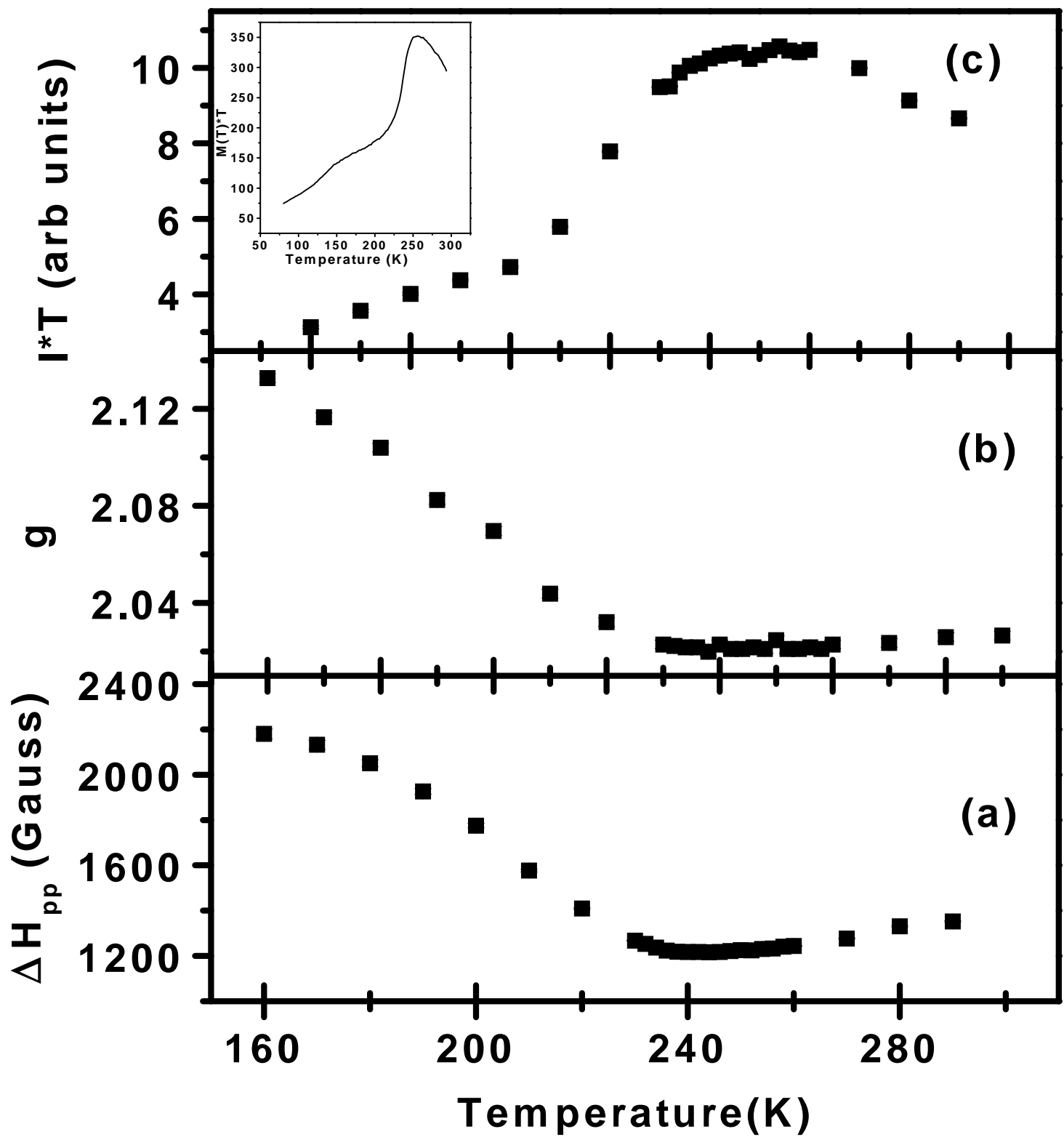
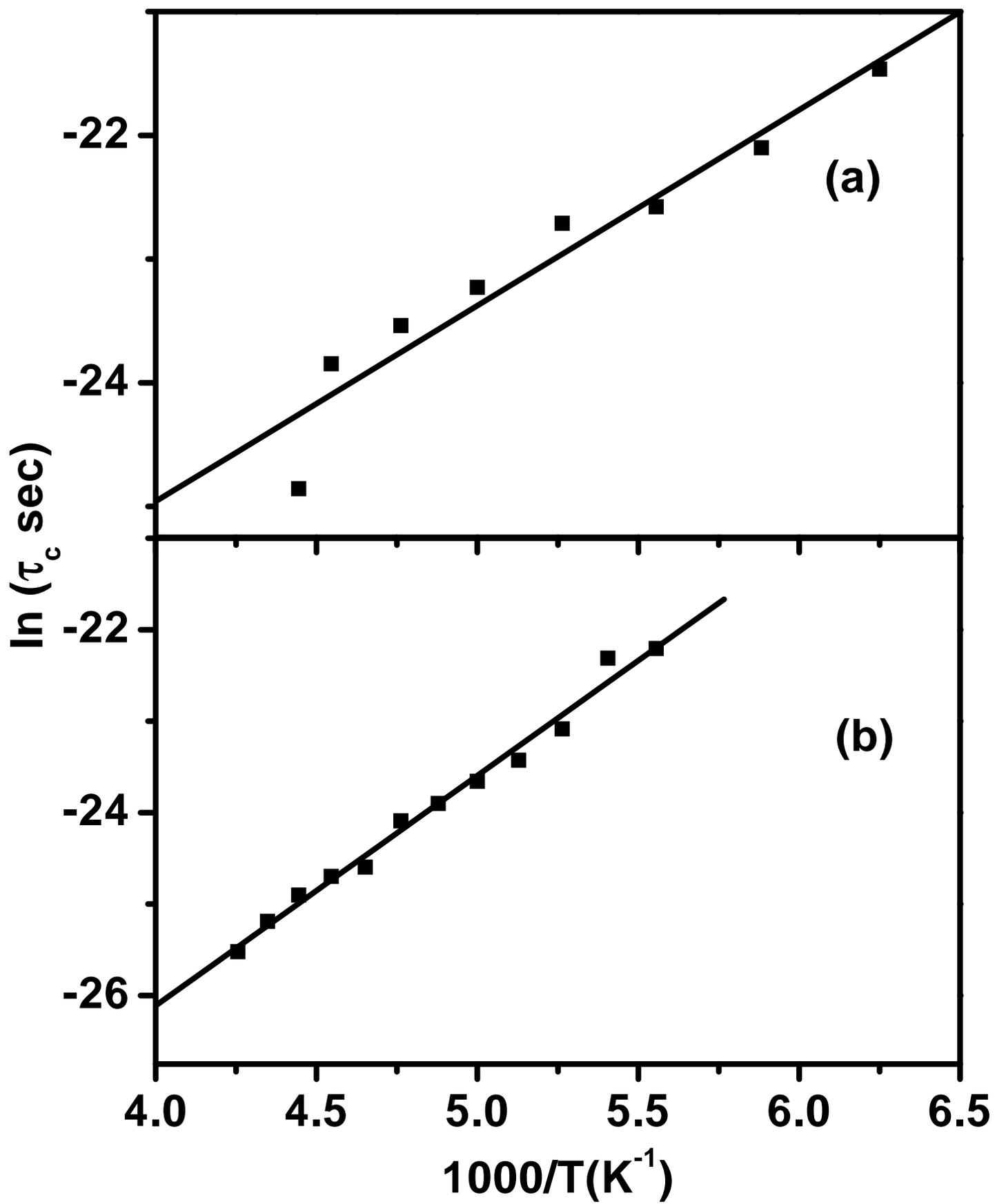
\end{document}